\documentstyle [12pt] {article}
\newcommand{\be}{\begin{equation}}
\newcommand{\ee}{\end{equation}}
\newcommand{\bea}{\begin{eqnarray}}
\newcommand{\eea}{\end{eqnarray}}
\newcommand{\ba}{\begin{array}}
\newcommand{\ea}{\end{array}}

\newcommand{\kb}{\bar{k}}
\newcommand{\pab}{\bar{\partial}}
\newcommand{\p}{\partial}
\newcommand{\zb}{\bar{z}}

\newcommand{\la}{\langle}
\newcommand{\ra}{\rangle}
\newcommand{\tb}{\bar{\tau}}
\newcommand{\ds}{\displaystyle}
\newcommand{\ap}{\alpha^\prime}
\newcommand{\ep}{\epsilon}
\newcommand{\nc}{\newcommand}

\nc{\btu}{\bigtriangleup}
\nc{\im}{{\rm Im}\ }
\nc{\re}{{\rm Re}\ }
\nc{\nn}{\nonumber}
\nc{\cd}{\cdot}
\nc{\lmf}{linear multiplet formalism }
\nc{\cmf}{chiral multiplet formalism }
\nc{\kae}{K\"ahler }
\begin{document}
\begin{titlepage}
\begin{flushright}MPI--Ph/93--19 \\ TUM--TH--153/93 \\
hep-th/9304055  \\
\ \\
March 1993
\end{flushright}
%

\vspace{0.8cm}
\begin{center}
{\Large \bf Dilaton, Antisymmetric Tensor and Gauge Fields in String
Effective Theories at the One--loop Level$^\ast$}   \\
\vspace{0.5cm}
{\large \bf P. Mayr}\ \ and\ \ {\large \bf S. Stieberger} \\
\vspace {0.5cm}

{ \em Physik Department} \\
{ \em Institut f\"{u}r Theoretische Physik} \\
{ \em Technische Universit\"at M\"unchen} \\
{ \em D--8046 Garching, W--Germany}
\begin{center}
{\rm  and}
\end{center}
{\em Max--Planck--Institut f\"ur Physik} \\
{ \em ---Werner--Heisenberg--Institut---}\\
{ \em P.O. Box 401212}\\
{ \em D--8000 M\"{u}nchen, W--Germany}
\end{center}
\normalsize
\begin{center}
\vspace{1cm}
{\bf ABSTRACT}
\end{center}
\begin{quote}
We investigate the dependence of the gauge couplings on the dilaton
field in string effective theories at the one--loop level. First we
resolve the discrepancies between statements based on symmetry
considerations and explicit calculations in string effective
theories on this subject. A calculation of the relevant one--loop
scattering amplitudes in string theory gives us further information
and allows us to derive the exact form of the corresponding effective
Lagrangian. In particular there is no dilaton dependent one--loop
correction to the holomorphic $f$--function arising from massive
string modes in the loop.
In addition we address the
coupling of the antisymmetric tensor field to the gauge bosons at
one--loop.
While the string S--matrix elements are not reproduced
using the usual supersymmetric Lagrangian with
the chiral superfield representation
for the dilaton field,
the analogue Lagrangian with the dilaton in a linear multiplet
naturally gives the correct answer.
\vspace{.3cm}
\hrule width 5.cm
{\scriptsize
\hspace{-.5cm} \noindent $^\ast$ Supported by the Deutsche
Forschungsgemeinschaft
and the EC under contract no SC1--CT92--0789.}
\normalsize
\end{quote}
\end{titlepage}
\setcounter{page}{1}
\voffset -2.5cm
\section{Introduction}
String theory is the only known theory consistent with quantized
gravity. Whether it is also consistent with the world where we all
live is a different and very difficult question. To tackle this
problem one tries to derive an effective Lagrangian, which describes
the low--energy limit of the string theory in a field--theoretical
language. One question of current interest are modifications of the
one--loop effective field theory through massive string modes running in the
loops. While the contributions of the massless modes are generated by
the field theory itself, those of the massive ones have to be plugged
in by hand. In general these stringy effects have to be determined
by explicit one--loop calculations
\cite{kap88,DKL91,ant13,ant14,dol92,li92}
directly in four--dimensional string theory.

In lucky cases there is yet another way to derive the string loop
corrections: if there are severe stringy symmetries and in addition
they are known. These symmetries can give significant restrictions on
the possible structure of the Lagrangian terms which can not be
understood from the point of field theory. This is because the
field theory as the low--energy limit has to
respect the symmetries of its stringy roots. This method was
partially successful in the case of a restricted class of
orbifold models for the moduli and gauge group dependent part of the
gauge couplings.  The discrete reparametrization symmetry
SL(2,{\bf{Z}}) of the moduli space was sufficient to infer
the exact form of the one--loop string corrections from the requirement
of anomaly freedom of the effective theory with respect to
SL(2,{\bf{Z}}) \cite{der91,car92}.
However already in this favourable case it was not
possible to determine the Lagrangian uniquely from symmetry
considerations \cite{der91}: to fix the coefficients of the two
independent counterterms, a Green--Schwarz term and a holomorphic
contribution from massive modes, again string calculations
are necessary \cite{ant13}. Furthermore it was shown in ref. \cite{thr92} that
for the majority of orbifold models the known symmetry group
is smaller than SL(2,{\bf{Z}}) and therefore this method
fails due to the lack of reliable information.

Another scalar field with a perturbatively flat potential
is the dilaton field, whose vev determines the gauge coupling
at tree--level \cite{wit85}.
Therefore it seems natural to ask for the
dependence of the one--loop corrections to the gauge coupling
on this field. Although there exist no appropriate quantum symmetries
comparable to the discrete
reparametrization symmetries of the moduli space, there
are classical symmetries \cite{bur86} which have quite strong implications.
Earlier investigations based on this symmetry have led to the
statement, that the one--loop piece of the gauge coupling is
independent of the dilaton vev \cite{nil86}. We will find in sect. 2
that this kind of statement is too rigorous as has been already
indicated by the results of explicit calculations in effective
theories, including dilaton dependent gauge couplings
\cite{lou91,car92,gai92}.
Nevertheless the above mentioned symmetries are sufficient
to determine the exact dependence of the one--loop gauge
couplings on the dilaton vev.

To confirm the symmetry arguments as well as to gain information
also about the couplings of the quantum fields, a string loop
calculation is desirable. Since string theory is an S--matrix
theory there are problems with on--shell singularities
in the computation of scattering amplitudes. We will use
Minahan's off--shell prescription \cite{min88} for the momentum variables
to handle this difficulty and present in the appendix the
derivation using the background--field method \cite{kap88,dol92} as an
alternative concept. Another issue is related to the fact that string
theory calculations provide always S-matrix elements: To obtain the 1PI
vertices generating effective Lagrangian one has to assign a
string correction to corrections to 1PI vertices in an appropriate
way. This step plays an important role for the translation
of string loop calculations into a field--theoretical language.

The outline of this paper is the following. In sect. 2 we discuss
the special role of the dilaton in string theory and its effective
theories. In sect. 3 we compute a three--point scattering amplitude
of two gauge bosons and a physical boson of the supergravity
multiplet on the torus. This can be a graviton, an antisymmetric
tensor field or the dilaton. In sect. 4 we discuss the obtained
results in terms of effective supersymmetric Lagrangians and
give our conclusions.
\section{The dilaton as the loop expansion parameter}
In this section we want to analyze earlier investigations on the
dilaton dependence of the gauge coupling at one--loop in string theory
based on symmetry considerations. The well--known statement arising
from this kind of reasoning is that the one--loop correction to the gauge
coupling is independent of the dilaton vev. This has to be contrasted
with the results of explicit calculations in
string effective field theories, which include a one--loop coupling
of the dilaton to the gauge bosons \cite{lou91,car92,gai92}.
A closer examination of the underlying assumptions present
in the symmetry arguments shows that there are no discrepancies to
worry about. Furthermore we will derive in this way the exact dilaton
dependence of the one--loop contribution to the gauge coupling
in string effective theories.

First let us consider the situation directly in string theory
\cite{sei86}.
The two--dimensional action of the superstring contains the expression
\be \label{marilyn}
\frac{1}{4\pi\ap}
\int d^2z \ \sqrt{h}
h^{\alpha\beta} \p_\alpha X^\mu \p_\beta X^\nu g_{\mu\nu}(X) +
\frac{1}{4\pi} \int d^2z \ \sqrt{h} R^{(2)} \phi(X) \ ,
\ee
where $h$ and $R^{(2)}$ are the two--dimensional metric and curvature,
respectively, and $\ap$ is the slope parameter.
$g_{\mu\nu}(X)$ corresponds to the background metric, while $\phi(X)$
is a scalar field.
The two--dimensional integration over the constant mode $\phi_0$ of $\phi$
yields
\be
\frac{1}{4\pi} \int d^2z \sqrt{h} R^{(2)} \phi_0 = \chi = 2(1-l) \phi_0 \ ,
\ee
where $\chi$ is the Euler characteristic and $l$ is the genus of the
world--sheet manifold. For this reason the contribution of higher genus
will be suppressed by a factor of $e^{-2l\phi_0}$ for each loop.
This is the same functional dependence on the genus as that of the
weight of a Riemannian surface in the path integral, $g^{2l}$,
where $g$ is a free parameter, the string coupling constant \cite{wei85}.
Therefore a change in the coupling constant can be compensated by a
redefinition of the field $\phi$ and in this way is determined by its vev.
To relate
the field $\phi$ (or parts of it) to the dilaton field one uses
the following form of the dilaton vertex operator in the case of
vanishing momentum \cite{sei86}:
\be \label{liane}
V_D^0(k \rightarrow 0) =  \frac{a}{4\pi\ap} \int d^2z \sqrt{h}
 \ h^{\alpha\beta} \p_\alpha X^\mu \p_\beta X^\nu \eta_{\mu\nu}
+ \frac{b}{4\pi} \int d^2z \sqrt{h} R^{(2)} \ .
\ee
Here $a$ and $b$ are some real numbers which can be determined
by demanding unitary and conformal invariance or alternatively
from the limiting field theory \cite{ber82,sug13}.
The term proportional to $R^{(2)}$ has been discussed extensively in the
literature \cite{dil00}.
However this issue will be of no importance for our discussion.
A small shift $\delta D$ of the dilaton vev can be taken into account
by the
insertion of (\ref{liane}) into each correlator:
\be \label{x}
\la V_1 ... V_N \ra \longrightarrow
\la V_1 ... V_N \ra +
\delta D \la \ V_1 ... V_N V_D^0(k\rightarrow 0) \ra \ .
\ee
In turn this can be absorbed to leading order in the associated
constant field $\delta D$
by modifying the background fields in (\ref{marilyn}) as
\bea \label{nina}
\phi(X) &\longrightarrow& \phi(X) + b \delta D \ , \nn \\
g_{\mu\nu}(X) &\longrightarrow& g_{\mu\nu}(X)(1+a \delta D) \ .
\eea
It follows that parts of the field $\phi$ can be identified
with the dilaton field and that also the normalization of the
space--time metric will be influenced by the vev of the dilaton field. A second
possibility would be to make the replacements
\bea \label{orphelia}
\phi(X) &\longrightarrow& \phi(X) + b \delta D \ , \nn \\
\ap &\longrightarrow& (1-a \delta D)\ap \ .
\eea
Note that (\ref{nina}) corresponds to a fixed string scale $M_S \sim
\alpha ^{\prime-1/2}$,
while (\ref{orphelia}) will leave unchanged the Planck mass $M_{pl}$
if $a$ and $b$ are
chosen correctly. Here the
Planck mass is defined as the coefficient of the Einstein term in the
low--energy Lagrangian or, equivalently, by the strength of the three
graviton interaction in string theory.

Since there is the relation
$2\kappa = (2\ap)^{\frac{d-2}{4}}g$
connecting the three parameters $\ap$, $g$ and $\kappa \sim M_{pl}^{-2}$
of the effective field theory in $d$ dimensions \cite{gro86,zou88}, there is
the possibility to form a dimensionless parameter proportional
to $ \alpha^{\prime (d-2)/4} \kappa^{-1}$. The above argumentation
recovers the well--known result that such a parameter will be determined
dynamically rather than representing a free input parameter
\cite{wit84}.

What are the implications for the dilaton dependence of a genuine
correlator? From eqs. (\ref{nina}) and (\ref{orphelia})
one can see that the
statement that a correlator is independent of the dilaton vev,
if the second term on the
r.h.s. of eq. (\ref{x}) vanishes relies heavily on the
field normalizations. In this context it is helpful to note that
the background field $g_{\mu\nu}$ appearing in eqs. (\ref{marilyn}) and
(\ref{nina}) is not the canonical normalized gravitational field
as it is used in the original determination of the effective Lagrangian
calculating string amplitudes \cite{gro86,zou88}. This point can be
easily seen also in the field--theoretical formulation.
 Eqs. (\ref{nina}) and (\ref{orphelia}) correspond to the following two
Lagrangians, respectively:
\bea
\label{petra}
\frac{{\cal L}}{e} &=&  \frac{\Phi^{-\frac{d-2}{4}}}{\kappa^2}
\left[-\frac{1}{2} R -\frac{1}{4}F^2 -\frac{3}{4}H^2
+ \ .\ .\ .\ \right] \ ,\\
\label{quendoline}
\frac{{\cal L}}{e} &=&   -\frac{1}{2\kappa^2} R - \frac{1}{4}
\frac{\Phi^{-1/2}}{g_d^2} F^2 - \frac{3}{4} \frac{\kappa^2}{g_d^4}
\Phi^{-1}H^2 + \ .\ .\ . \ \ \ \   .
\eea
We use the standard notations for the fields \cite{gro86}
except for $\Phi$
which is defined to be $\Phi = e^{\frac{4\kappa}{\sqrt{d-2}}D}$, where $D$
is the $d$--dimensional dilaton field.
In eq. (\ref{petra}) the space--time metric has been rescaled with
respect to (\ref{quendoline}) as $g_{\mu\nu} \rightarrow
\Phi^{-1/2} g_{\mu\nu}$
and the string scale $M_S \equiv g_d/\kappa$ has been set to 1.
Apparently the factor in front of (\ref{petra}) can be
identified with the loop expansion parameter \cite{sei85}.
Therefore in this special field basis the one--loop gauge coupling
should be independent of the dilaton VEV in any dimension.
This Lagrangian corresponds to
the case (\ref{nina}) with a dilaton independent normalization of the
background metric $g_{\mu\nu}$. This choice causes a dilaton
dependent
coefficient of the Einstein term in the effective Lagrangian.
In the more physical formulation of (\ref{quendoline})
where $M_{pl}$ is fixed, the terms of the Lagrangian have an individual
dependence on $\Phi$ and the loop expansion parameter is in general
different from the coefficient of the $F^2$ term. Therefore one
can no longer infer a dilaton independence of the gauge coupling
at one--loop and we see again that the usual reasoning based on symmetry
considerations
is only valid for a certain choice of field normalizations
corresponding to (\ref{petra}).

It is important that the dependence of the gauge coupling on $\Phi$ in
(\ref{quendoline})
is different for any number of dimensions. This reflects in some
sense the singular behavior of a Yang--Mills theory in $d$ dimensions.
In d=4 the special situation arises that the $\Phi$ dependence of the
coefficients of the $F^2$ terms in (\ref{petra}) and (\ref{quendoline}) agree.
Nevertheless a simple cancellation of
couplings will not be sufficient to ensure the dilaton independence
of the one--loop contribution to the gauge kinetic term:
the Lagrangian
(\ref{quendoline}) describes a field theory with a dilaton dependent
physical cutoff which should show up in an one--loop effective
gauge coupling. In summary we can state the following result:
The dilaton dependence of the gauge coupling at one--loop in the
physical field basis with $M_{pl} \neq M_{pl}\ (\la D\ra)$ is entirely due to
the
dilaton dependence of the cutoff $\sim M_S$ ( in four dimensions ).

It is useful to notice the precise assertion of the above argumentation.
First note that it is valid for all one--loop
corrections to the gauge coupling, regardless whether they arise as
a one--loop term of the holomorphic $f$--function of supergravity
\cite{sug13} or from non--holomorphic terms \cite{DKL91}. On the other hand it
is restricted to the constant part of the dilaton field, that is, its
vev.

A similar discussion can be carried on for the symmetry arguments
of ref. \cite{nil86}, valid for the four--dimensional effective theory
in the limit of infinitely small compactification radius.
In this paper the (at least) classical symmetries
of the effective field theory \cite{bur86}
arising from the degeneration of the string
vacuum were used to narrow down the possible form of the effective
Lagrangian. Two kinds of symmetries were considered there: First two
classical scale invariances of the theory representing the flat
directions in the effective potential of the dilaton and the
universal modulus field related to the overall radius of
the compactification manifold. Secondly two axial $U(1)$ symmetries
given by constant shifts of the supersymmetric pseudoscalar partners of the
above mentioned fields. This axial symmetries are valid to all
orders of perturbation theory and can be traced back to
the invariance of the
field strength of an antisymmetric tensor field.

Two important
results were derived in ref. \cite{nil86}
from the presence of these symmetries:
using the scale invariances together with the input that the
loop expansion parameter is given by $S/T$, it was shown that
the one--loop kinetic term of the gauge fields should be independent
of the $S$ field. Here $S$ ($T$) is the chiral superfield containing the
four--dimensional dilaton (overall modulus) field. In fact this
result was stated only for the holomorphic $f$--function, but the assumption
of holomorphicity was not yet used at this point. In addition
the axial $U(1)$ symmetries together with holomorphicity were used
to prove that the gauge coupling receives no contributions from
higher than one--loop. This remarkable conclusion is not transferable to the
now well--known non--holomorphic one--loop corrections, but is only
valid for the gauge coupling of the effective action defined in the
spirit of ref.
\cite{wil74} \footnote{See also ref. \cite{shi86}.}. It is clear that the field
basis with dilaton
dependent $M_{pl}$ was used throughout ref. \cite{nil86}.
\section{The two gauge boson -- dilaton amplitude in string theory}
In the previous section we derived the precise dilaton dependence
of the one--loop corrections to the gauge coupling using symmetry
consideration. To confirm our results as well as to get information
also about the coupling of the non--constant part of the dilaton
field to the gauge bosons at one--loop, we calculate the
CP even part of the two gauge boson -- dilaton coupling
at one--loop in string theory.

We will consider in general a three--point function on the torus involving
two gauge bosons and a physical boson from the supergravity multiplet.
This provides a useful check of the correct normalizations relative
to existing calculations. Furthermore we want to get information
about the coupling of the antisymmetric tensor field to the
gauge bosons. Our
conventions and normalizations are given in appendix A. We
will compute a scattering amplitude and compare the result
with that of a background--field calculation in appendix B.

The CP even part of the
three--point function at one--loop is given by
\bea
{\cal A} &=& \frac{1}{2}
\sum_{even \ s}(-)^{(s_1+s_2)} \nonumber \\
&\times&\int _{\tau \in \Gamma}
 \frac{d^2 \tau}{{\rm Im} \ \tau} Z(\tau,\tb , s) \int \prod _{i = 1}
^{3} d^2 z_i
\langle V^G_0 (z_1,\zb_1) V^A_0(z_2,\zb_2)
V^A_0(z_3,\zb_3)
\rangle_s \ ,
\eea
where $s_1$,$s_2$ label the three CP even spin structures $(s_1,s_2)$ =
(0,0), (0,1), (1,0). A zero (one) corresponds to a NS (R) boundary
condition. $\tau = \tau _1 + i \tau _2$ is the single Teichm\"{u}ller
parameter of the worldsheet torus and $\Gamma$ its fundamental domain
$\Gamma = \{\tau \ | \ |\tau_1| < 1/2 \ ; \ | \tau| > 1\}$.
 The partition function for the heterotic string is
$Z(\tau,\tb,s) \equiv Tr_{s_1} [(-)^{s_2F} q^{H-1/2}
q^{\bar{H}-1}]$ with $q = e^{2\pi i \tau}$ and
$V_0^A$,$V_0^G$ are the zero--picture vertex operators for the gauge boson
and the supergravity tensor, respectively:
\bea \label{fabienne}
V^A_0(z_1,\zb_1) &=& \frac{4g}{\pi}
\epsilon _{1\mu} J^a_{\bar{1}}(\partial X^\mu_1 + i \psi _1 ^\mu
k_1 \cdot \psi _1) e^{ik_1X_1} \nn \ , \\
V^G_0(z_1,\zb_1) &=& \frac{8\kappa}{\pi} \epsilon _{\mu \nu} ^{(1)}
\bar{\partial} X_1^\mu
(\partial X^\nu_1 + i\psi _1 ^\nu k_1 \cdot \psi _1) e^{ik_1X_1} \ .
\eea
Here we use the notation $f(z_i) \equiv f_i$, $f(\zb_i) \equiv
f_{\bar{i}}$ \footnote{We will denote the dependence
$f(z_i,\zb_i)$
again with $f_i$ if no confusion is possible.}. $g$ is
the dimensionless string coupling parameter and $\kappa$ the gravitational
coupling.
We will set
$\alpha^\prime = \frac{1}{2} $
in most of the following  unless
an explicit representation is useful to distinguish
 terms of different
order in $\alpha^\prime$. The full dependence can be recovered by
simple dimensional considerations.

Because of supersymmetry, summing over the spin structures will
give a zero result unless there are right--handed fermions in the
correlation function. However, since the fermions come in
normal--ordered pairs, there must actually be four fermions
to get a non--vanishing result. Therefore, performing all possible
contractions which fulfill this condition yields the following
expression for the amplitude:
\bea
\ds \label{ariane}
{\cal A} &=& (2\alpha^\prime) \frac{4\kappa g^2}{\pi^3}
\sum_{even \ s}(-)^{(s_1+s_2)}
\int _{\tau \in \Gamma} \frac{d^2 \tau}{{\rm Im} \ \tau}
Z(\tau,\tb , s) \int \prod _{i = 1} ^{3} d^2 z_i
\prod_{i<j} \left| \chi_{ij} \right| ^{k_ik_j\ap} \langle J_{\bar{2}}
J_{\bar{3}} \rangle \nonumber \\
&\times&\Bigg\{ \Bigg[ \bigg( \sum_{i<j}
(\ep_i \cdot k_j  \ep_j \cdot k_i -
\ep_i \cdot \ep_j  k_i \cdot k_j)   G^2_{ij}(s) \bigg)
\bigg( \sum _{p \neq i,j  \atop q \neq p}
\ep_p \cdot k_q  \p_{pq} {\rm ln}  |\chi_{pq}|^2 \bigg)\nonumber \\
&& \; \: \; \; \;-  4 t(1,2,3)
\ G_{12}(s) G_{13}(s) G_{23}(s) \Bigg] \times
\left[ \ep^*_1 \cdot k_2  \p_{\bar{1}\bar{2}} {\rm ln}  |\chi_{\bar{1}
\bar{2}}|
^2 +
\ep^*_1 \cdot k_3  \p_{\bar{1}\bar{3}} {\rm ln}
|\chi_{\bar{1}\bar{3}}|^2
\right] \nonumber \\
&& \; \; \; - \ \frac{16}{2\ap} V \bigg( \sum_{i<j}
(\ep_i \cdot k_j \ \ep_j \cdot k_i -
\ep_i \cdot \ep_j \ k_i \cdot k_j) G^2_{ij}(s)\bigg)
\bigg( \sum_{p \neq i,j} \ep_1^* \cdot \ep_p \bigg)
\Bigg\} \ .
\eea
$G(s)$ is the spin structure dependent fermionic two--point
function and $t(1,2,3)$ is the
kinematical factor of (\ref{uli})
(for further details see appendix A).
In (\ref{ariane}) the polarization
tensor $\ep^{(1)} _{\mu \nu}$ has been
replaced by $\ep_{1\mu}^* \ep_{1\nu}^{}$ for calculational reasons.
Eq. (\ref{ariane})
differs from that of ref. \cite{min88} by an additional
term proportional to $V \equiv -\frac{\pi}{4 \im \tau}$ in
(\ref{ariane}) which stems from the
contraction of the fields $\p X_1$ and $\pab X_{1}$
in the dilaton vertex operator. In general a
contraction of the fields $\p X$ and $\pab X$ yields
\be \label{kaltenberg-hell}
\langle \p X_1 \pab X_2 \rangle = \p_1 \p_{\bar{2}}
\left( - \frac{1}{4} {\rm ln} \ |\chi_{12}|^2 \right)
= \frac{\pi}{4} \delta^2(z_{12}) - \frac{\pi}{4 \im
\tau} \ ,
\ee
where the second term is the contribution coming from
the zero modes. While it is obvious that the
$\delta$--function  disappears if normal--ordered fields
are considered, more involved arguments are necessary
to show that this part of the two--point function
does not contribute even in the more general case of
unordered fields. Actually there are several different ways to
derive this result \cite{min88,pan87,gre87}.

The next step is to extract from (\ref{ariane}) the terms of
lowest order in $\ap$. Apart from the obvious terms
proportional to $V$ there arise additional ones from
the integrations over the $z$'s in regions where
some of the vertex operators are close together on
the world--sheet. The resulting singularities can be treated
along the line of Minahan's
off--shell prescription for the kinematical
variables, which respects both, modular and conformal
invariance \cite{min88}. The crucial point is that it is by no means
fixed a priori, whether the momentum integral is performed
before or after the $z$--integrations. This allows one
to choose the kinematical variables which respect only the
constraints from conformal and modular invariance but not
from momentum conservation during parts of the
calculation. The correctness of this procedure was
confirmed in ref. \cite{ber89}, where the authors show that the
same result is obtained by factorising a four--point
amplitude.

The additional terms arise in the following way: the
correlation functions behave for $z_{ij}
\rightarrow 0$ as
\be
G_{ij} \rightarrow \frac{1}{4z_{ij}} \ , \hspace{1cm}
\p_{ij} \ln |\chi_{ij}|^2 \rightarrow \frac{1}{z_{ij}}  \ , \hspace{1cm}
|\chi_{ij}|^2 \rightarrow |z_{ij}|^2 \ .
\ee
Integration over those terms in (\ref{ariane}) which provide a pole
in $z_{ij}$ yields, by analytic continuation
\be
\int_{|z_{ij}| < \ep} d^2z_{ij} \frac{1}{|z_{ij}|^2} |z_{ij}|^{\ap k_i \cdot
k_j} = \frac{2\pi}{\ap k_i \cdot k_j} \ .
\ee
Therefore this mechanism provides a source for
new ${\cal O}(\alpha^{\prime 0})$ terms\footnote{Of
course there is an additional
dependence on $\ap$ implicit in $\kappa$.}
with poles in $k_i
\cdot k_j$, which cancel similar factors in the kinematical
factors of (\ref{ariane}). The resulting expression of
${\cal O}(\alpha^{\prime 0})$ reads
\bea \label{barbara}
{\cal A}=&& \frac{16\kappa g^2}{\pi^2}
\sum_{even \ s}(-)^{(s_1+s_2)}
\int _{\tau \in \Gamma}
 \frac{d^2 \tau}{{\rm Im} \ \tau} Z(\tau,\tb , s)
\int d^2z_{23} \langle J_{\bar{2}} J_{\bar{3}} \rangle
\nn \\
&\times&\Big\{\im \tau \ {\rm tr} (\ep^{(1)})
\ep_2 \cdot k_3 \ep_3 \cdot k_2 G_{23}^2(s)
\\
&&+ 2 ( k_3 \cdot \ep^{(1)}_{as} \cdot \ep_3 \ep_2 \cdot k_1
 + k_2 \cdot \ep^{(1)}_{as} \cdot \ep_2
\ep_3 \cdot k_1) \int d^2z  G^2(z,s) \Big\} \ , \nn
\eea
where
\be
\ep^{(1)}_{as} \equiv \frac{1}{2} (\ep_1-\ep_1^T) \ .\nonumber
\ee
Note that there is no contribution arising from
$z_{23} \rightarrow 0$. In addition also the region
where all vertex operators come close together does not
contribute because of supersymmetry.

The two terms inside the curly brackets of (\ref{barbara}) form
separately gauge invariant objects. In fact two of the
long distance contributions proportional to $V$ together
with the pole terms\footnote{
The distinction in pole and long distance terms is only
convention: by partial integration one can change this
'classification'.} are gauge invariant as well as the
single term involving ${\rm tr} (\ep^{(1)})$.
 This is not
surprising because exactly this structure asserts gauge
invariance for the three different types of polarization
tensors corresponding to a graviton, an antisymmetric
tensor or a dilaton.

In (\ref{barbara}) the only spin structure dependent factors are
the fermionic correlation functions. Therefore we can
rewrite them using (\ref{veronique}). After this replacement we
perform the remaining $z$--integrations and splitting the
partition function in space--time and internal part we
get
\be \label{Daniela}
{\cal A} = {\cal K} \frac{\kappa g^2}{8\pi^2}
\int _{\tau \in \Gamma} \frac{d^2\tau}{\im \tau}
{\cal B}_a \ ,
\ee
where
\bea \label{Claudia}
{\cal K} &=& -\frac{1}{2} {\rm tr} (\ep^{(1)})
\ep_2 \cdot k_3 \ep_3 \cdot k_2
+ k_3 \cdot \ep^{(1)}_{as} \cdot \ep_3 \ep_2 \cdot k_1
+ k_2 \cdot \ep^{(1)}_{as} \cdot \ep_2 \ep_3 \cdot k_1 \ ,
\\
\label{test}
{\cal B}_a &=& \sum_{even \ s}(-)^{(s_1+s_2)}
2 |\eta(\tau)|^{-4} \frac{-1}{2\pi i} \frac{d}{d \tau}
Z_\psi (s) \\
&& \times
{\rm Tr}_s\ \Big[ \big(Q_a^2 + \frac{k_a}{4\pi
\im \tau} \big) (-)^{s_2F}q^{H-3/8} \bar{q}^{\bar{H}-11/12}
\Big] _{(c,\bar{c})= (9,22)} \ .\nonumber
\eea
{}From (\ref{Claudia}) it is clear that there is a
zero result for the case where the polarization
tensor is that of a graviton. In the case of the
antisymmetric tensor the result agrees with
that of ref. \cite{li92} after the
identification
\be
k_3 \cdot \ep^{(1)}_{as} \cdot \ep_3 \ep_2 \cdot k_1
+ k_2 \cdot \ep^{(1)}_{as} \cdot \ep_2 \ep_3 \cdot k_1
\longrightarrow \frac{1}{2} A_\mu \p_\nu A_\rho
(\p^\mu B^{\nu \rho} + \p^\nu B^{\rho \mu}
+ \p^\rho B^{\mu\nu}) \ ,
\ee
in spite of the fact that the derivation as well as the interpretation
therein seems to be incorrect\footnote{Compare with the discussion
of eq.(\ref{kaltenberg-hell}).}.
In fact there is also a non--vanishing result if $\ep^{(1)}$ is chosen to
describe a dilaton. In this case the polarization
tensor is given by
\be \label{f2783}
\epsilon ^{(1)} _{\mu \nu} = \frac{1}{\sqrt{d-2}}
(\eta _{\mu \nu} - k_{\mu} \kb _{\nu} - k _{\nu} \kb _{\mu}) \ ,
\ee
where we have defined a vector $\kb$ such that $\kb ^2 = 0$ and $k \cdot \kb
= 1$. Thus
\be
{\rm tr} (\ep^{(1)}) = \sqrt{d-2} = \sqrt{2} \ ,
\ee
and (\ref{Daniela}) becomes
\be \label{Elen}
{\cal A} = -\sqrt{2} (\ep_2 \cdot k_3 \ep_3 \cdot k_2 - \ep_2 \cdot \ep_3 k_2
\cdot k_3)
\frac{\kappa g^2}{16 \pi ^2}
\int_{\tau \in \Gamma} \frac{d^2 \tau}{\im \tau}
{\cal B}_a \ ,
\ee
where we have reinstated a kinematical term
which was dropped during the calculations.
Eq. (\ref{Elen}) gives the following
term to an one--loop S-matrix element:
\be \label{helena}
\frac{1}{4}\tilde{\btu}^D_a F^a_{\mu\nu}F_a^{\mu\nu}D \ ,
\ee
where
\be \label{spezi}
\tilde{\btu}_a^D = \sqrt{2} \kappa \btu_a^g = \sqrt{2} \kappa
\left( \frac{1}{16\pi^2}
\int_{\tau \in \Gamma} \frac{d^2 \tau}{\im \tau}
{\cal B}_a \right) \ .
\ee
The tilde on $\tilde{\btu}_a^D$ denotes that it represents a correction
to an S-matrix element rather than to an 1PI vertex. Furthermore
$\btu_a^g$ is the one--loop correction to the
gauge coupling derived in \cite{kap88}
and the gauge fields have been rescaled as
$A \rightarrow gA$.

If the internal part of the theory is
given by an orbifold compactification \cite{orb00} it is
possible to make more concrete statements. In this case the
moduli dependence
of the one--loop correction to the gauge coupling $\btu^g_a$ has been discussed
first in ref. \cite{DKL91}.
In particular it has been argued there that only moduli
corresponding to a plane which is left unchanged by a subgroup of the
orbifold group can enter the corrections. The orbifold sectors
which are generated by elements of this subgroup leave intact two
space--time supersymmetries and therefore build up an N=2 sector of the
theory. For the restricted class of orbifold compactifications where
the six--dimensional torus splits into a direct sum, $T_6 = T_2 \oplus T_4$,
and the fixed plane lies in $T_2$,
explicit expressions for $\btu^g_a$ have been obtained with the help
of an auxiliary model involving a two--dimensional torus compactification.
The functional dependence of $\btu^g_a$ on a generic modulus $T$
has been found to be an automorphic function of the duality group $SL(2,{\bf
Z})$:
\be
\btu^g_a\sim \ln [|\eta(T)|^4(T + \bar{T})] \ ,
\ee
where $\eta(T)$ is the Dedekind function.
Recently the more general case where the internal torus is generated
by a genuine six--dimensional lattice,
was presented in ref. \cite{thr92}.
It was shown there that the so--called threshold function $\btu^g_a$
is no longer an
automorphic function of the modular group $SL(2,{\bf Z})$ but only of
a certain subgroup of $SL(2,{\bf Z})$.
\section{Connection to field theory}
We turn now to the discussion of the previous string calculation in terms
of an effective Lagrangian. In the field theory two different
types of diagrams provide corrections to the S--matrix element
$\tilde{\btu}^D_a$ of eq.(\ref{spezi}): wave function
renormalizations $\delta\Pi_a$ and $\delta\Pi_D$ in an external
gauge and dilaton leg, respectively, as shown in fig. 1 as well as genuine 1PI
corrections to the three--particle vertex shown in fig. 2:
\be \label{do}
\tilde{\btu}^D_{a} = (p_1^\nu p_2^\mu - g^{\mu\nu} p_1 \cdot p_2)
\left[ \{ g_{a,D}^{-2}\}^{1-loop} + (2 \delta\Pi_a + \delta\Pi_D)
\{ g_{a,D}^{-2}\}^{tree} \right] \ ,
\ee
where $ \{ g_{a,D}^{-2}\}^{1-loop}$ and $\{ g_{a,D}^{-2}\}^{tree}$ denote
the effective linear coupling of the dilaton to gauge bosons at one--loop
and tree--level, respectively . In contrast to the
analogous relation for the moduli dependent correction to the gauge
coupling \cite{DKL91}, where the second term is absent since
$\{ g_{a,T}^{-2}\}^{tree} = 0$ ($T$ denotes the moduli field), the
loops in the external legs now contribute because the dilaton couples
to the gauge bosons at tree--level:
\be \label{just}
\{ g_{a}^{-2}\}^{tree} = e^{-\sqrt{2}\kappa D} \ .
\ee
To assign the string correction $\tilde{\btu}^D_a$ to the two terms in
eq. (\ref{do}) let us focus on the gauge group dependent part $\sim Q_a^2$
in eq. (\ref{test}). If we consider only differences between the corrections
to the gauge couplings of different gauge groups, as it was introduced
in ref. \cite{DKL91} to get rid of the modular non--invariant regulator
terms of the correlator of the Kac--Moody currents, the term $\sim \delta
\Pi_D$ drops. The expression for the wave--function renormalization
factor $\delta\Pi_a$ can be extracted from ref. \cite{kap88} and multiplying
it by the tree--level coupling $ \{ g_{a,D}^{-2}\}^{tree}$ of
eq. (\ref{just}) we reproduce exactly the string correction $\tilde{\btu}^D_a$.
In conclusion there is no one--loop correction to the effective
dilaton--gauge boson coupling $\{ g_{a,D}^{-2}\}^{1-loop}$ at all.

On the other side we have to consider the corrections calculated in field
theory as well as the presence of the dilaton dependent cutoff $\sim M_S$. From
the
discussion of section 2 we expect that the complete effective Lagrangian
should depend on the dilaton through $M_S$.
Obviously this will be a subject involving only massless string modes.
Because of the lack
of information about the explicit form of the effective theory in
general four--dimensional string models, we have to restrict
our investigation from now on to the case of orbifold compactifications.
The usual representation
of the dilaton field in the supergravity theory is to use a chiral
superfield $S$ with a lowest scalar component given by
Re $s = e^{-\sqrt{2}\kappa D}$. There
are two sources of a coupling of the superfield $S$ to the gauge superfields
at one--loop. The first is due to the tree--level coupling contained in the
supersymmetric completion of a Lagrangian term corresponding to
eq.(\ref{just}). The second one is the presence of $S$ in the \kae connection,
which in turn appears in the covariant derivatives of the fermionic fields
\cite{der91,lou91}.
The \kae connection is related to the U(1) rotations
on the fermions which are necessary to make the supersymmetric Lagrangian
invariant under \kae transformations. If this U(1) symmetry is gauged
at the superfield level, the arising theory is called \kae  superspace
\cite{bin87}.
The coupling of the \kae connection and on--shell gauge bosons
generated by triangle graphs with fermions running in the loop can be
represented by a non--local contribution to the
effective Lagrangian \cite{sup83,lou91,der91,car92}:
\bea \label{woodstock}
{\cal L}_{nl} &=& \frac{1}{4} \frac{1}{(4\pi)^2} \int d^4 \theta \sum_a
(W^aW^a) \frac{D^2}{\Box} \frac{1}{2} \left(c_{adj}^a - \sum_R c_R^a \right)
\kappa^2 K \ + {\rm h.c.}  \nn \\
&=& - \frac{1}{4} \frac{1}{(4\pi)^2} \int d^4 \theta \sum_a
(W^aW^a) \frac{D^2}{\Box} \frac{1}{2} \left(c_{adj}^a - \sum_R c_R^a \right)
\\
&&\times
\left(\ln (S+\bar{S}) + \sum_{I} \ln(T+\bar{T})\right)  + {\rm h.c.} \ ,\nn
\eea
where we have only indicated the pure dilaton and
moduli dependent part of the \kae potential.
$W^a$ is the chiral superfield containing the Yang--Mills field
strength $F_{\mu\nu}^a$ of the gauge group labeled by $a$,
$c_{adj}^a$ ($c_R^a$) is the quadratic
Casimir operator in the adjoint (relevant matter) representation and the sum
runs over the matter representations $R$. In general it is also possible
for a chiral superfield to couple to the fermions through the covariantization
of the derivatives with respect to sigma--model general coordinate
transformations \cite{der91,lou91}. Since this coupling is absent for the
superfield $S$ in string inspired supergravity theories because there
are simply no charged fields with a dilaton dependent metric at
the tree--level,
it is irrelevant for this discussion. Nevertheless we have yet to take
into account the above mentioned contribution due to the tree--level
coupling of $S$ to the gauge bosons. In ref. \cite{lou91} it has been
shown that this term gives an expression similar to that of eq.
(\ref{woodstock}) with a group theoretical factor exactly of the size
to complete the coefficient in eq. (\ref{woodstock}) to that of the full
$\beta$--function.
In this way one gets
the following algebraic equation for the effective gauge couplings,
valid to all orders and written in general as a function of a
chiral superfield $\Phi$:
\bea
g_a^{-2}(\Phi,\bar{\Phi},p) &=& \re f_a(\Phi) + \frac{b_a}{8\pi^2}
\ln \ \frac{M_{pl}}{p} - \frac{\kappa^2}{16\pi^2}
\left(c_{adj}^a - \sum_R c_R^a \right) K(\Phi,\bar{\Phi}) \nn \\
&&+ \frac{c_{adj}^a}{8\pi^2} \ln \ g_a^{-2}(\Phi,\bar{\Phi},p) -
\frac{1}{8\pi^2} \sum_R c^a_R \ln \ \det \ Z_R(\Phi,\bar{\Phi},p) \ ,
\eea
where $p$ is the renormalization scale, $b_a$ is the group theoretical
coefficient of the $\beta$--function $\beta_a = b_ag_a^3/16\pi$ and
$Z_R$ is the field dependent \kae metric for the matter fields in
the representation $R$. If we insert the tree--level \kae potential
for $S$ as well as the tree--level gauge coupling and furthermore
replace Re $S$ by its vev, this term has the effect
to shift the scale at which the couplings begin to run from $M_{pl}$
to $M_S$:
\be
b_a \ln\frac{M^2_{pl}}{\mu^2} - b_a \ln \la \re S \ra = b_a \ln
\frac{M_S^2}{\mu^2} \ ,
\ee
due to the relation $M_{pl}/M_S = \la \re s \ra ^{1/2}$. Here $s$ denotes
the lowest component of the superfield $S$.
This is the result expected from the discussion of the previous sections.
The fact that the parity even piece of the non--local Lagrangian
(\ref{woodstock})
can be interpreted as a supersymmetric cutoff was first pointed out in ref.
\cite{gai91}.

There are two further points worth mentioning: first we emphasize that in
general there could be yet a holomorphic piece in the one--loop correction
to the gauge coupling, which  contains the field $S$. This would correspond
to a one--loop piece of the $f$--function. From the previous results
it is clear that this term is absent in the string effective theory,
$\re f_a^{1-loop}(S)= 0$.
Secondly it is nice to record the consistent normalization of the
$T$ field. At first sight it looks conspicuous that the $T$ field
in the field theory has to be normalized in units of $M_{pl}$, while
that appearing in the string calculations is  apparently normalized
in units of $M_S$. But it is easy to convince oneself that in the
string theory an expansion of the field--dependent gauge coupling
in powers of $T$ includes a factor $g$ for each field $T$.
Therefore the precise expression for the $T$ dependent functions
appearing in the string calculations of refs. \cite{DKL91,thr92} is
obtained by the replacements $T[M_S] \rightarrow
gT[M_S] = T[M_{pl}]$ in the expressions given therein, in agreement
with the field--theoretical description. Here we have used
$g = \la \re s \ra ^{-1/2}$.

Now let us take a look at the coupling of the antisymmetric tensor field
$B_{\mu\nu}$ to the gauge bosons at one--loop, which is related to that
of the dilaton field by supersymmetry. From eq. (\ref{Daniela}) we find
$\tilde{\btu}_a^B = - \btu_a^g$. Again there is a tree--level coupling
due to the presence of the Chern--Simons term in the definition
of the field strength $H$:
\be
H_{\mu\nu\rho}= \p_{[\rho} B_{\mu\nu]} - \frac{\kappa}{4}
{\rm tr} \ \left( A_{[\rho} F_{\mu\nu]} - \frac{g}{3}
A_{[\mu} A_{[\nu}A_{\rho]]} \right) \ .
\ee
Subtracting from $\tilde{\btu}_a^B$ the contribution of the loops
in the external legs, as we did before for the case of the dilaton field,
we find again a zero result. That is, the one--loop S--matrix element
in string theory is entirely reproduced by the diagram shown in fig. 3.

In an unnoticed way we have run into trouble. On one hand we have shown that
there are no one--loop corrections to the three particle vertices involving
two gauge bosons and a dilaton or an antisymmetric tensor field in the
effective string
theory. On the other hand there is the expression in eq. (\ref{woodstock})
which shifts nicely the scale $M_{pl}$ to $M_S$ after it is
supplemented with the contribution
related to the tree--level coupling of $S$ to the gauge bosons
and if the field $(S+\bar{S})$ is replaced by its vev.
However, if we make an expansion around this vev,
the same term induces linear couplings of two gauge bosons and a dilaton
or an antisymmetric tensor field:
\bea \label{bauchweh}
{\cal L}_{nl} &=&- \frac{1}{4} \frac{1}{(4\pi)^2} \int d^2 \theta \sum_a
(W^aW^a)  \frac{\bar{D}^2 D^2}{\Box} \frac{1}{2}\ b_a \nn \\
&&\times \left[ \la \ln (S+\bar{S}) \ra + \frac{1}{\la S+\bar{S} \ra}
(\tilde{S}+\tilde{\bar{S}}) + ... \ \ \right] + {\rm h.c.} \\
&=& \frac{1}{16} \frac{1}{(4\pi)^2}
\sum_a b_a \nn \\
&& \times\left[ F^aF^a \left( \la \ln (S+\bar{S}) \ra
+ \frac{1}{\la S+\bar{S} \ra} (\tilde{S}+\tilde{\bar{S}})
\right) - F^a \tilde{F}^a
 \frac{1}{\la S+\bar{S} \ra} (\tilde{S}-\tilde{\bar{S}}) + ...\  \right], \nn
\eea
where the tilde on $\tilde{S}$ identifies the quantum field. This is
in apparent contradiction with the situation we have found in string theory.
We will argue that this problem appears due to an inconsistent formulation
of the effective theory in terms of the chiral superfield $S$.

In fact it is known that the four--dimensional N=1 supersymmetric
effective target space superstring theories correspond to so--called
new--minimal supergravities with a linear multiplet containing the
dilaton field \cite{cec87,eva88}.
The general coupling of a linear multiplet to the supergravity--matter
system and Yang--Mills fields was derived in \cite{lin87} and recently
the complete supersymmetric action was constructed \cite{ada92}.

It is well--known that a supergravity theory including a linear multiplet
can be related to a theory with only chiral matter fields via a duality
transformation \cite{lin87}. Such a transformation will connect an
effective theory in the \lmf to another one in the \cmf in a classical
sense. Therefore one should not start to compute quantum corrections
of the two theories related in this way and compare the results with
each other\footnote{We thank J.P. Derendinger for an explanation
on this issue.}. This is exactly what happens in the above mentioned
problem: we have compared the quantum corrections of a theory in the
\cmf obtained from a tree--level duality transformation to that of
the one--loop corrections in a theory formulated with a linear multiplet.
Indeed if we replace the \kae potential of the dilaton superfield $S$,
$K(S,\bar{S}) = - \kappa^{-2}\ln (S+\bar{S})$ by that of the linear
multiplet containing the dilaton, $K(L) = \kappa^{-2}\ln L$ we get
\bea \label{it}
{\cal L}_{nl} &=&  \frac{1}{4} \frac{1}{(4\pi)^2} \int d^2 \theta \sum_a
(W^aW^a)  \frac{\bar{D}^2 D^2}{\Box} \frac{1}{2}\ b_a \nn \\
&&\times \left[ \la \ln L \ra + \frac{1}{\la L \ra}
L + ... \ \ \right] + {\rm h.c.} \\
&=& - \frac{1}{8} \frac{1}{(4\pi)^2}
\sum_a b_a \left[ F^aF^a \left( \la \ln L \ra
+ ... \ \right) - F^a \tilde{F}^a
\left( 0 + ... \right) \right] \nn \ .
\eea
Here we have used $D^2L=\bar{D}^2L=0+ ...$ , where the dots represent
terms involving field strengths arising from the modified linearity
conditions
\bea \label{tja}
(D_{\dot{\alpha}}D^{\dot{\alpha}} - 8 R) \ L &=& 2k {\rm tr}\ (W^aW^a)\nn \ ,\\
(D^{\alpha}D_{\alpha} - 8 R^\dagger) \ L &=& 2k {\rm tr}\ (\bar{W}^a\bar{W}^a)
\ ,
\eea
where $R$ is one of the torsion superfields of supergravity. We have
written only the coupling to the gauge field strength on the r.h.s. of
(\ref{tja}) which corresponds to adding the Yang--Mills Chern--Simons form
to the field strength of the antisymmetric tensor. In general also
Chern--Simons terms for the Lorentz and U(1) group are necessary to get
an anomaly free theory.
{}From eq. (\ref{it}) we can see that there are no linear couplings of dilaton
or antisymmetric tensor fields to the gauge bosons in the linear
multiplet formalism.

At first sight this seems to be a strange result because the \kae
connection in the \lmf contains a term \cite{ada92}
\be
V_\mu^L = ... \ -\frac{i}{8} \frac{1}{L} \ep_{\mu\nu\rho\lambda}\p^\nu
B^{\rho\lambda} \ .
\ee
Therefore $B_{\mu\nu}$ indeed couples to the fermions analogue to
the coupling of the corresponding pseudoscalar $a = \im s$ in the
\cmf:
\be
V_\mu^C= ... \ - \frac{i}{2} \frac{1}{S+\bar{S}} \p_\mu a \ .
\ee
We have seen in eq. (\ref{bauchweh}) that this coupling causes a non--vanishing
one--loop correction to the two gauge boson -- pseudoscalar vertex.

To analyze the different outcome of the calculation of the triangle
diagrams in the two formalisms we write down the action of the operator
$\Box^{-1}\bar{D}^2D^2$ on $S$ and $L$, respectively:
\bea \label{raf}
\Box^{-1}\bar{D}^2D^2 S |_{\theta = \bar{\theta} = 0} &=&
\Box^{-1}\ 16\ \Box (\re s + i a ) \ , \nn \\
\Box^{-1}\bar{D}^2D^2 L |_{\theta = \bar{\theta} = 0} &=&
\Box^{-1}\ 4i\ \p_\mu A^\mu \ ,
\eea
where $A_\mu = \ep_{\mu\nu\rho\lambda}\p^\nu B^{\rho\lambda}$. In the
first line of
eq. (\ref{raf}) the operator $\Box^{-1}\Box$ is ill--defined on--shell,
but since it is off--shell everywhere equal to one,
it is one also on--shell,
by analyticity. The contrary applies to the second line
of eq. (\ref{raf}). First note that the dilaton dependent term has already
dropped out because of an algebraic identity. Furthermore the expression
$\ep_{\mu\nu\rho\lambda}p^\mu p^\nu B^{\rho\lambda}$ vanishes for
any (complex) momentum $p_\mu$. Therefore, by analyticity,
$\Box^{-1} \p_\mu A^\mu$ has to be zero also on--shell. If we replace
$A_\mu$ by the duality transformed expression $\p_\mu a $, the algebraic
zero reflecting the gauge invariance of the antisymmetric tensor field
strength turns into a weaker on--shell zero of the Laplacian acting on
the pseudoscalar field $a$. This fact is in the heart of the difference between
the non--local Lagrangians of eqs. (\ref{bauchweh}) and (\ref{it}).

Of course this effect can be seen also at the component level
and is not restricted to a supersymmetric model. The explicit
expression for the triangle graph with two vector gauge bosons
$A_\sigma^1$, $A_\rho^2$ and one axial gauge boson $A_3^\mu$ was
given long ago in ref. \cite{ros63}:
\bea
R_{\sigma \rho \mu}(k_1,k_2) &=& A_3 \left( k_1 \cdot k_2\ k_1^\tau
\ep_{\tau\sigma\rho\mu} - k_1 \cdot k_2\ k_2 ^\tau \ep_{\tau\sigma\rho\mu}
+ k_{1\rho} k_1^\xi k_2^\tau \ep_{\xi\tau\sigma\mu}
-k_{2\sigma} k_1^\xi k_2^\tau \ep_{\xi\tau\rho\mu} \right)\nn \\
&&+ A_4 \left( k_2^2k_1^\tau \ep_{\tau\sigma\rho\mu} - k_1^2k_2^\tau
\ep_{\tau\sigma\rho\mu} + k_{2\rho} k_1^\xi k_2^\tau
\ep_{\xi\tau\sigma\mu} - k_{1\sigma} k_1^\xi k_2^\tau \ep_{\xi\tau\rho\mu}
\right) \ ,
\eea
where $k_1$ and $k_2$ are the momenta of the vector gauge bosons and
$A_3$ and $A_4$ are some integrals over Feynman parameters. The ambiguity
of the linearly divergent diagram has been fixed by the requirement of
gauge invariance in the vector channel.
With the handicap of massless fermions in the loop it is straightforward to
show the following relations for on--shell vector bosons $A_\sigma^1$ and
$A_\rho^2$:
\be\ba{lll}
R_{\sigma \rho \mu}A^\mu &= 0 &{\rm for} \ A^\mu = \ep^{\mu\nu\lambda\kappa}
\p_\nu B_{\lambda\kappa} \\
 R_{\sigma \rho \mu}A^\mu &= 8\pi^2k_1^\xi k_2^\tau
\ep_{\xi \tau \sigma \rho} a &{\rm for}\ A^\mu = \p^\mu a\ . \\
\ea
\ee
In agreement with the previous result in the supersymmetric formulation
there is only a non--vanishing coupling if the axial vector boson
is written in terms of a pseudoscalar.

In conclusion we have shown that the one--loop gauge coupling in string
effective theories depends on the dilaton vev in a way determined by the
dilaton dependence of the cutoff scale $M_S$. This result can be
derived by symmetry considerations and applies to the holomorphic
as well as to the non--holomorphic piece of the one--loop correction.
The identification of the dilaton dependence as a
field--theoretical effect shows that it is entirely due to massless fields
and, in particular, that there is no contribution from massive modes providing
a dilaton dependent one--loop correction to the holomorphic $f$--function.
Calculating a three point string amplitude involving two gauge
bosons and one dilaton we have shown that there is no one--loop correction to
the three particle vertex in the 1PI vertices generating effective
Lagrangian. The same statement applies to the coupling of the
antisymmetric tensor field to the gauge bosons. This result is not reproduced
by the usual non--local term in the effective Lagrangian
in the formalism, where
the dilaton is in a chiral multiplet. On the other hand the analogue effective
theory written in terms of a linear multiplet is in full agreement with
string theory.

\section*{Acknowledgements}
We would like to thank J.P. Derendinger, J. Louis and H.P. Nilles
for priceless discussions. We also return thanks to D. Jungnickel,
A. Klemm and S. Theisen for support.
\renewcommand{\theequation}{\thesection.\arabic{equation}}
\newcommand{\sect}[1]{ \section{#1} \setcounter{equation}{0} }
\section*{Appendix}
\appendix
\sect{Conventions and normalizations}
\rm %
The overall normalizations in the operator formalism can
be fixed by comparison with results from a path integral
calculation. In our conventions the correlation function
for the bosonic fields (with $\ap = \frac{1}{2}$) reads
\be \label{wanda}
\langle X^\mu(z_1,\zb_1) X^\nu(z_2, \zb_2) \rangle
= - \frac{1}{4} \eta ^{\mu \nu} {\rm ln}
|\chi(z_{12},\tau)|^2 \ ,
\ee
where $z_{12} = z_1 - z_2 $ and
\be \chi_{ij} \equiv
\chi (z_{ij}) = 2 \pi \exp\left[- \pi
\frac{({\rm Im} \ z_{ij})^2}
{{\rm Im} \ \tau} \right] \left|
\frac{\vartheta _1(z_{ij}|\tau)}{\vartheta_1^\prime(0|\tau)}
\right| \ .
\ee
$\vartheta_1 (z|\tau)$ is one of the Riemann
theta--functions. Correlation functions of the fields
$\p X$, $\pab X$ can be obtained by simply taking
derivatives of (\ref{wanda}) except for $\langle \p X \pab X \rangle$
as discussed in the text. They obey the identities
\be \label{null}
\int d^2 z\ \p_z \ln |\chi(z)| = 0 \ ,\hspace{1cm}
\int d^2 z\ \p^2_z \ln |\chi(z)| = 0 \ .
\ee
The fermionic correlation function is
\be G_{12}(s) \equiv
\la\psi _1 \psi _2\ra _s = \frac{1}{4}
\frac{ \vartheta
_\alpha (z_{12} | \tau)
\vartheta _1^\prime (0 |\tau)}{\vartheta _1 (z_{12} | \tau)
\vartheta
_\alpha (0|\tau)} \ ,
\ee
where $\alpha$ = 2,3,4 for ($s_1,s_2$) = (1,0), (0,0), (0,1), respectively.
Furthermore we have defined
$\vartheta ^\prime _1 (0|\tau) = \p _z \vartheta _1 (z|\tau) |_{z = 0}$.
If there is no additional
spin structure dependence but that of $(G_{12}(s))^2$ only the spin
structure dependent terms inside
this correlation function will survive and we can
replace
\be \label{veronique}
\left(4G_{12}(s)\right)^2 =
- 2
\p^2_{12} \ln \chi_{12} - \hat{G}_2(\tau) -
 e_{\alpha-1}(\tau)
\longrightarrow
 4\pi i \frac{d}{d\tau} \ln \frac{\vartheta_\alpha
(0|\tau)}{\eta(\tau)} \ ,
\ee
where $\hat{G}_2(\tau)$ is the Eisenstein function of modular
weight two and the
$e_{\alpha-1}(\tau)$ are defined in terms of the
Weierstrass function\footnote{For the usual conventions
see \cite{mag48}}.

The correlation function for the Kac--Moody currents is
given by
\be \label{yuhu}
\la J^a_{\bar{1}}  J^b_{\bar{2}}\ra = \frac{1}{16}
\big[ - k_a \delta ^{ab} \p ^2
_{\bar{1}\bar{2}}
\ln \bar{\Theta} _1
(\bar{z}_{12} \mid \tb) - 4 \pi^2 Q_a Q_b \big] \ ,
\ee
where $k_a$ is the level of the Kac--Moody algebra
and the $Q_a$'s are the charges
of the propagating states.

The four--dimensional free plus ghost part of the
partition function is given by
\bea
Z_\psi(\tau,\tb,s) &=& \frac{\vartheta_\alpha(0|\tau)}
{\eta(\tau)} \nn \ , \\
Z_{B}(\tau,\tb) &=&
\frac{1}{|\eta(\tau)|^4}
\frac{1}{(2\pi^2 \im \tau)^2}  \ ,
\eea
for the fermions and bosons, respectively. In addition there is a contribution
$Z_{int}$ from the internal (c,$\bar{c}$) = (9,22)
superconformal theory.

The contraction of six fermions inside a
correlator yields the kinematical factor
\bea \label{uli}
t(1,2,3)&=&\ep_1 \cd k_3 \ep_2 \cd k_1 \ep_3 \cd k_2 - \ep_1 \cd k_2 \ep_2 \cd
k_3 \ep_3 \cd k_1 \nn \\
&&+\ \ep_1 \cd \ep_2
[\ep_3 \cd k_1 k_2 \cd k_3 - \ep_3 \cd k_2 k_1 \cd k_3] \nn \\
&&+\ \ep_1 \cd \ep_3 [ \ep_2 \cd k_3 k_1 \cd k_2
- \ep_2 \cd k_1 k_2 \cd k_3] \nn \\
&&+\ \ep_2 \cd \ep_3 [ \ep_1 \cd k_2 k_1 \cd k_3
- \ep_1 \cd k_3 k_1 \cd k_2 ] \ \ .
\eea
\sect{Background-field calculation}
%
The vertex operators for the background fields can be
obtained from (\ref{fabienne}) by the replacements
\be \ep_\mu e^{ikX} \longrightarrow A_\mu(X) \ ,  \hspace{2cm}
\ep_{\mu\nu} e^{ikX} \longrightarrow G_{\mu\nu}(X) \ .
\ee
The fields $A_\mu(X)$ and $G_{\mu\nu}(X)$ are solutions of the
classical equations of motions. Note that the $X$--dependence
of the background fields has to be taken into account when the
contractions on the world--sheet are performed. Keeping only the
first terms in an expansion gives
\bea \label{susi}
V_0^A(z_1,\zb_1) &=& \frac{4g}{\pi} J^a_{\bar{1}}
(A_{1\mu} \p X_1^\mu + \p_\nu A_{1\mu} X_1^\nu \p X_1^\mu
 + \psi_1^\mu
\psi_1^\nu \p_\nu A_{1\mu}) \nn \ , \\
V_0^G(z_1,\zb_1) &=& \frac{8\kappa}{\pi} \pab X_1^\mu
(\p X_1^\nu G_{\mu\nu}^{(1)} + \p X_1^\nu X^\lambda \p_\lambda G_{\mu\nu}^{(1)}
+ \psi ^\nu \psi ^\lambda \p_\lambda G_{\mu\nu}^{(1)} ) \ .
\eea
Only three terms in the correlator $\la V_0^GV_0^AV_0^A \ra$
yield non--vanishing contributions of ${\cal O}(\alpha^{\prime 0})$. They are
\bea \label{gina}
&&\la \pab X_1^\mu \p X_1^\nu \ra
\la \psi_2^\gamma \psi_2 ^\rho \psi_3 ^\alpha \psi_3 ^\beta \ra \la J_{\bar{2}}
J_{\bar{3}} \ra  G_{\mu\nu}^{(1)}
\p _\gamma A_{2\rho} \p_\alpha A_{2\beta} \ , \nn \\
&& \la \pab X_1^\mu \p X_3^\alpha \ra
\la \psi_1^\lambda \psi_1 ^\nu \psi_2 ^\gamma \psi_2 ^\rho \ra \la J_{\bar{2}}
J_{\bar{3}} \ra A_{3\alpha}
\p_\lambda G_{\mu\nu}^{(1)} \p_\gamma A_{2\rho} \ , \\
&&
\la \pab X_1^\mu \p X_2^\rho \ra
\la \psi_1^\lambda \psi_1 ^\nu \psi_3 ^\alpha \psi_3 ^\beta \ra \la J_{\bar{2}}
J_{\bar{3}} \ra A_{2\rho}
\p_\lambda G_{\mu\nu}^{(1)} \p_\alpha A_{3\beta} \ . \nn
\eea
The further calculation is quite similar to the case involving the usual
vertex operators. The second and third term in (\ref{gina}) supply
the amplitude where $G_{\mu\nu}$ refers to the antisymmetric tensor.
If we restrict our attention to the dilaton field, (\ref{gina}) results in
\be
+ \frac{\kappa}{4}{\rm tr} G_{\mu\nu}^{(1)} F_{\alpha\beta}^aF_a^{\alpha\beta}
\left( \frac{g^2}{16\pi^2}
\int_{\tau \in \Gamma} \frac{d^2 \tau}{\im \tau}
{\cal B}_a \right) \ ,
\ee
in agreement with (\ref{helena}).
\newpage
\footnotesize
\renewcommand{\baselinestretch}{0.1}

\vspace{2cm}
\section*{Figure Captions}
\normalsize
\renewcommand{\baselinestretch}{1.3}
\baselineskip15pt
{\bf Figure 1:} One--loop contributions to the three--particle S--matrix
element with one dilaton and two gauge bosons arising from
loops in the external legs. Wavy lines denote gauge bosons while
the dashed line represents the dilaton.\\ \\
{\bf Figure 2:} One--loop correction to the same S--matrix element associated
to the 1PI three--particle vertex.
The wavy--solid line denotes any massless field.\\ \\
{\bf Figure 3:} Contributions to the one--loop S--matrix element with
one antisymmetric tensor field and two gauge bosons. The double line
represents the antisymmetric tensor field.
\end{document}